\documentclass[pre,twocolumn,aps]{revtex4}
\newcommand{\bec}[1]{\mbox{\boldmath $ #1$}}
\usepackage{graphicx}
\usepackage{bm}
\newcommand{\Onecol} {\begin{widetext} \onecolumngrid} 
\newcommand{\Twocol} {\end{widetext} \twocolumngrid} 

\newcommand{\BE}[1]{\begin{equation}\label{eq:#1}} 
\newcommand{\EE}{\end{equation}}
\newcommand{\BEA}[1]{\begin{eqnarray} \label{eq:#1}} 
\newcommand{\BEa}{\begin{eqnarray*}} 
\newcommand{\EEa}{\end{eqnarray*}} 
 \def\br {\\ \nonumber} 
\newcommand{\B}[1]{{\bm{#1}}}
 \def\r{\B {r}} 
\def\<{\left \langle} \def\>{\right\rangle}
\newcommand{\ok}{(\omega,\B k)}
\newcommand{\tr}{(t,\B r)}

\def\Fbox#1{\vskip1ex\hbox to 8.5cm{\hfil\fboxsep0.3cm\fbox{%
\parbox{8.0cm}{#1}}\hfil}\vskip1ex\noindent} 
\def\K41{\fboxrule0.2ex\fbox{\large\text{K41}}}
\def\Fbox#1{\vskip1ex\hbox to 8.5cm{\hfil\fboxsep0.3cm\fbox{%
\parbox{8.0cm}{#1}}\hfil}\vskip1ex\noindent} 
\begin{document}

\title{Clustering of Aerosols in Atmospheric Turbulent Flow}
\author{T. Elperin$^{a}$}
\email{elperin@bgu.ac.il}
\author{N. Kleeorin$^a$}
\author{M. A. Liberman$^b$}
\author{V.~S.~L'vov$^c$}
\author{I. Rogachevskii$^a$}
\affiliation{$^a$The Pearlstone Center for Aeronautical Engineering
 Studies, Department of Mechanical Engineering,
 Ben-Gurion University of the Negev,
 P. O. Box 653, Beer-Sheva
84105, Israel\\
$^b$Department of Physics, Uppsala University, Box 530, SE-751
21, Uppsala, Sweden\\
 $^c$Department of Chemical Physics,
 The Weizmann Institute of Science,
 Rehovot 76100, Israel}

\date{\today}

\begin{abstract}
A mechanism of formation of small-scale inhomogeneities in spatial
distributions of aerosols and droplets associated with clustering
instability in the atmospheric turbulent flow is discussed. The
particle clustering is a consequence of a spontaneous breakdown of
their homogeneous space distribution due to the clustering
instability, and is caused by a combined effect of the particle
inertia and a finite correlation time of the turbulent velocity
field. In this paper a theoretical approach proposed in Phys. Rev. E
{\bf 66}, 036302 (2002) is further developed and applied to
investigate the mechanisms of formation of small-scale aerosol
inhomogeneities in the atmospheric turbulent flow. The theory of the
particle clustering instability is extended to the case when the
particle Stokes time is larger than the Kolmogorov time scale, but
is much smaller than the correlation time at the integral scale of
turbulence. We determined the criterion of the clustering
instability for the Stokes number larger than 1. We discussed
applications of the analyzed effects to the dynamics of aerosols and
droplets in the atmospheric turbulent flow.

\bigskip

\end{abstract}
\maketitle

\section{Introduction}

It is known that turbulence enhances mixing (see, e.g.,
\cite{MY75,C80,PS83,MC90,S96,B97,W00,S01,BH03}). However, numerical
simulations, laboratory experiments and observations in the
atmospheric turbulence revealed formation of long-living
inhomogeneities in spatial distribution of aerosols and droplets in
turbulent fluid flows (see, e.g.,
\cite{WM93,KM93,EF94,FK94,MC96,SC97,KS01,AC02,S03,CK04,CC05,AG06}).
The origin of these inhomogeneities is not always clear but their
influence on the mixing can be hardly overestimated.

It is hypothesized that the atmospheric turbulence enhances the rate
of droplet collisions  (see, e.g., \cite{S03,PKK97,VY00,PK00,RC00}).
In particular, the turbulence causes formation of small-scale
droplet inhomogeneities, and it also increases the relative droplet
velocity. In addition, the turbulence affects the hydrodynamic
droplet interaction. The latter increases the rate of droplet
collisions. These effects are of a great importance for
understanding of rain formation in atmospheric clouds. In
particular, these effects can cause the droplet spectrum broadening
and acceleration of raindrop formation \cite{S03,VY00}. Note that
clouds are known as zones of enhanced turbulence. The preferential
concentration of inertial particles (particle clustering) was
recently studied in numerical simulations in \cite{HC01,B03,BL04}.
The formation of network-like regions of high particle number
density was found in \cite{BL04} in high resolution direct numerical
simulations of inertial particles in a two-dimensional turbulence.

The goal of this study is to analyze the particle-fluid interaction
leading to the formation of strong inhomogeneities of aerosol
distribution due to a \emph{particle clustering instability}. The
particle clustering instability is a consequence of a spontaneous
breakdown of their homogeneous space distribution. As a result, at
the nonlinear stage of the clustering instability, the local density
of aerosols may rise by orders of magnitude and strongly increase
the probability of particle-particle collisions.

It was suggested in \cite{EKA96,EKB96,EK98} that the main cause of
the particle clustering instability is their inertia: the particles
inside the turbulent eddies are carried out to the boundary regions
between the eddies by the inertial forces. This mechanism of the
preferential concentration acts in all scales of turbulence,
increasing toward small scales. Later, this was contested in
\cite{EKA00,EKA01} using the so-called "Kraichnan model" \cite{K68}
of turbulent advection by the delta-correlated in time random
velocity field, whereby the clustering instability did not occur.

However, it was shown in \cite{EK02} that accounting for a finite
correlation time of the fluid velocity field results in the
clustering instability of inertial particles. Note that the particle
inertia results in the compressibility of particle velocity field.
The effects of compressibility of the velocity field on formation of
small-scale inhomogeneities in spatial distribution of particles
were first discussed in \cite{K94,EK95}. In this study a theoretical
approach proposed in \cite{EK02} is further developed and applied to
investigate the mechanisms of formation of small-scale aerosol
inhomogeneities in the atmospheric turbulent flow. In particular, we
extended the theory of particle clustering instability to the case
when the particle Stokes time is larger than the Kolmogorov time
scale, but is much smaller than the correlation time at the integral
scale of turbulence.

Remarkably, the particle inertia also results in formation of the
large-scale inhomogeneities in the vicinity of the temperature
inversion layers due to excitation of the large-scale instability
(see \cite{EKA96,EKA00,EKRA00}). This effect is caused by additional
non-diffusive turbulent flux of particles in the vicinity of the
temperature inversion (phenomenon of turbulent thermal diffusion).
The characteristic time of excitation of the large-scale instability
of concentration distribution of aerosols varies in the range from
0.3 to 3 hours depending on the particle size and parameters of the
atmospheric turbulent boundary layer and the temperature inversion
layer. The phenomenon of turbulent thermal diffusion was recently
detected experimentally using two very different turbulent flows
created by oscillating grids turbulence generator
\cite{BEE04,EE04,EE06} and multi-fan turbulence generator
\cite{EEK06} for stably and unstably stratified fluid flows.

The paper is organized as follows. In Sec.~\ref{s:clust} we present
governing equations and a qualitative analysis of the clustering
instability that causes formation of particle clusters in a
turbulent flow. In Sec.~\ref{s:crit-val} we estimate the scalings of
the  particle velocity in the turbulent fluid for the case when the
particle Stokes time is much larger than the Kolmogorov time scale,
but is much smaller than the correlation time at the integral scale
of turbulence. In Sec.~\ref{moments} we perform a quantitative
analysis for the clustering instability of the second moment of
particle number density for ${\rm St} > 1$, where ${\rm St}$ is the
Stokes number. This allows us to generalize the criterion of the
clustering instability obtained in \cite{EK02}. Finally, in
Sec.~\ref{s:disc} we overview the nonlinear effects which lead to
saturation of the clustering instability and determine the particle
number density in the cluster. In Sec.~\ref{s:disc} we perform
numerical estimates for the dynamics of aerosols and droplets in
atmospheric turbulent flow. The conclusions are drawn in
Sec.~\ref{s:concl}. The detail analysis of the scalings of the
particle velocity in the turbulent fluid is given in Appendix A. The
detail analysis of the clustering instability of the inertial
particles is given in Appendix B.

\section{Governing equations and qualitative analysis of particle clustering}
\label{s:clust}

To analyze dynamics of particles we  use the standard continuous
media approximation, introducing the number density field $n(t,
\mathbf{r}) $ of spherical particles with radius $a$. The particles
are advected by an incompressible turbulent velocity field
${\mathbf{u}}(t,\mathbf{r}) $. The particle material density
$\rho_p$ is much larger than the density $\rho$ of the ambient
fluid. For inertial particles their velocity
${\mathbf{v}}(t,{\mathbf{r}})\neq{\mathbf{u}}(t,{\mathbf{r}}) $ due
to the particle inertia and ${\rm div} \,
{\mathbf{v}}(t,{\mathbf{r}}) \not = 0$  (see \cite{MC86,M87}).
Therefore, the compressibility of the particle velocity field
${\mathbf{v}}(t,{\mathbf{r}}) $ must be taken into account. The
growth rate of the clustering instability, $\gamma$, is proportional
to $\langle | {\rm div} \, {\mathbf{v}}(t,{\mathbf{r}})|^{2}\rangle
$  (see \cite{EKA96,EKB96,EK95}), where $\langle \cdot \rangle $
denotes ensemble average.

Let $\Theta(t,{\mathbf{r}}) $ be the deviation of the instantaneous
particle number density $n(t, \mathbf{r}) $ from its uniform mean
value $N \equiv \langle n \rangle $: $\, \Theta(t,{\mathbf{r}})=n(t,
\mathbf{r})-N$. The pair correlation function of
$\Theta(t,{\mathbf{r}}) $ is defined as
$\Phi({\mathbf{R}},{\mathbf{r}},t)\equiv\langle
\Theta(t,{\mathbf{r+R}})\Theta(t,{\mathbf{r}})\rangle$. For the sake
of simplicity we will consider only a spatially homogeneous,
isotropic case when $\Phi({\mathbf{R}},{\mathbf{r}},t) $ depends
only on the separation distance $R$ and time $t$, i.e.,
$\Phi(t,{\mathbf{R}},{\mathbf{r}})=\Phi(t,R)$. Clearly, a large
increase of $\Phi(t,R)$ above the level of $N^{2} $ can lead to a
strong grows in the frequency of the particle collisions.

In the analytical treatment of the problem we use the standard
equation for $n(t, \mathbf{r}) $:
\begin{equation}
\frac{\partial n(t, \mathbf{r})}{\partial
t}+\mathbf{\nabla}\cdot[n(t,
\mathbf{r}){\mathbf{v}}(t,{\mathbf{r}})]=D\triangle n(t,
\mathbf{r}), \label{5}
\end{equation}
where D is the coefficient of molecular (Brownian) diffusion. We
study the case of small yet finite molecular diffusion $D$ of
particles. The equation for $\Theta(t,\mathbf{r}) $ follows from
Eq.~(\ref{5}):
\begin{eqnarray}\label{6}
\frac{\partial \Theta(t, \mathbf{r})}{\partial
t}&+&[{\mathbf{v}}(t,{\mathbf{r}})\cdot\nabla]\Theta(t, \mathbf{r})
= -\Theta(t, \mathbf{r}) \, {\rm div} \,
{\mathbf{v}}(t,{\mathbf{r}})
\nonumber\\
&& +D\triangle \Theta(t, \mathbf{r})\ .
\end{eqnarray}
To study the clustering instability we use Eq.~(\ref{6}) without the
source term $\propto - N \, {\rm div}\,{\mathbf{v}} $, describing
the effect of an external source of fluctuations. Particle
clustering can also occur due to this source of fluctuations of
particle number density. Such fluctuations were studied in
\cite{B03,BL04,BF01}. In the present study we considered the
particle clustering due to the clustering instability. Particle
clustering caused by the self-excitation of fluctuations of particle
number density (the clustering instability) is much stronger than
that due to the source of fluctuations of particle number density.

One can use Eq.~(\ref{6}) to derive equation for $\Phi(t,R) $ by
averaging the equation for $\Theta(t,{\mathbf{r+R}})
\Theta(t,{\mathbf{r}})$ over statistics of the turbulent velocity
field ${\mathbf{v}}(t,{\mathbf{r}}) $. In general this procedure is
quite involved even for simple models of the advecting velocity
fields (see, e.g., \cite{EK02}). Nevertheless, the qualitative
understanding of the underlying physics of the clustering
instability, leading to both, the exponential growth of $\Phi(t,R) $
and its nonlinear saturation, can be elucidated by a more simple and
transparent analysis.

Let us consider turbulent flow with large Reynolds numbers,
${\mathcal{R}}e \gg 1 $. Therefore, the characteristic scale $L $ of
energy injection (outer scale) is much larger than the length of the
dissipation scales (\emph{viscous scale} $\eta $) $L\gg\eta $. In
the so-called \emph{inertial interval} of scales, where $L>r>\eta $,
the statistics of turbulence within the Kolmogorov theory is
governed by the only dimensional parameter, $\varepsilon $, the rate
of the turbulent energy dissipation. Then, the velocity $u(r)$ of
turbulent motion at the characteristic scale $r $ (referred below as
r-eddies) may be found by the dimensional reasoning:
$u(r)\approx(\varepsilon r)^{1/3}$ (see, e.g.,
\cite{MY75,LL87,F95}). Similarly, the turnover time of $r $-eddies,
$\tau(r)$, which is of the order of their life time, may be
estimated as $\tau(r)\approx r/u(r)\approx
\varepsilon^{-1/3}r^{2/3}$.

To elucidate the clustering instability let us consider a cluster of
particles with a characteristic scale $\ell $ moving with the
velocity ${\mathbf{V}}_{\rm cl}(t) $. The scale $\ell $ is a
parameter which governs the growth rate of the clustering
instability, $\gamma $. It sets the bounds for two distinct
intervals of scales: $L>r>\ell $ and $\ell>r>\eta $. Note also that
we cannot consider scales which are smaller than the size of
particles. Large $r $-eddies with $r>\ell $ sweep the $\ell
$-cluster as a whole and determine the value of ${\mathbf{V}}_{\rm
cl}(t) $. This results in the diffusion of the clusters, and
eventually affects their distribution in a turbulent flow.

On the other hand, the particles inside the turbulent eddies are
carried out to the boundary regions between the eddies by the
inertial forces. This mechanism of the preferential concentration
acts in all scales of turbulence, increasing toward small scales.
The role of small eddies is multi-fold. First, they lead to the
turbulent diffusion of the particles within the scale of a cluster
size. Second, due to the particle inertia they tend to accumulate
particles in the regions with small vorticity, which leads to the
preferential concentration of the particles. Third, the particle
inertia also causes a transport of fluctuations of particle number
density from smaller scales to larger scales, i.e., in regions with
larger turbulent diffusion. The latter can decrease the growth rate
of the clustering instability. Therefore, the clustering is
determined by the competition between these three processes.

Let us introduce a dimensionless parameter $\sigma_{\rm v} $, a
\emph{degree of compressibility} of the velocity field of particles,
${\mathbf{v}}(t,{\mathbf{r}}) $, defined by
\begin{equation}
\sigma_{\rm v} \equiv {\langle[{\rm div} \, {\mathbf{v}}]^{2}\rangle
\over \langle|\nabla \times{\mathbf{v}}|^{2}\rangle} \ . \label{19}
\end{equation}
This parameter may be of the order of 1 (see \cite{EKA00}). One of
the reasons for the clustering instability is the particle inertia
which results in the parameter $\sigma_{\rm v} \not=0$. The particle
response time is given by
\begin{equation}
\tau_p=\frac{m_p}{6\pi \, \nu\, \rho \, a}=\frac{2\rho_p \,
a^{2}}{9\rho \, \nu}\,, \label{31}
\end{equation}
and the particle mass $m_p $ is $ m_p=(4\pi/3) \, a^{3} \, \rho_p$.
The ratio of the inertial time scale of the particles (the Stokes
time scale $\tau_p $) and the turnover time of $\eta $-eddies in the
Kolmogorov micro-scale $\tau(\eta)=\eta/u(\eta)=\eta^{2}/\nu$, is of
primary importance, where $u(\eta)$ is the characteristic velocity
of $\eta $-scale eddies. The ratio of the time-scales $\tau_p$ and
$\tau(\eta)$ is the Stokes number:
\begin{equation}
{\rm St} \equiv \frac{\tau_p}{\tau(\eta)}=\frac{2\rho_p \, a^{2}}
{9\rho \, \eta^{2}}\ . \label{35}
\end{equation}
For $\tau_p \ll \tau(\eta)$ all particles are almost fully involved
in turbulent motion, and one concludes that
$u(t,{\mathbf{r}})\approx v(t,{\mathbf{r}}) $ and $v(\ell)\approx
u(\ell) $. The compressibility parameter $\sigma_{\rm v}$ of
particle velocity field for ${\rm St} \ll 1$ is given by:
\begin{equation}
\sigma_{\rm v} \sim \left(\frac{2\rho_p}
{9\rho}\right)^{2}\left(\frac{a}{\eta}\right)^{4} = {\rm St}^2 \, .
\label{46}
\end{equation}
(see \cite{EK02,EKA00}). For small Stokes number, the clustering
instability has been investigated in \cite{EK02}. The characteristic
scale of the most unstable clusters of small particles is of the
order of Kolmogorov micro-scale of turbulence, $\eta $. The
characteristic growth rate of the clustering instability is of the
order of the turnover frequency of $\eta $-eddies, $1/\tau(\eta) $
(see \cite{EK02}). In the present study we extend the theory of
particle clustering instability to the case ${\rm St} > 1$, i.e.,
when the particle Stokes time is larger than the Kolmogorov time
scale, but is much smaller than the correlation time at the integral
scale of turbulence. We may expect that for ${\rm St} > 1$ the
compressibility parameter $\sigma_{\rm v} $ of particle velocity
field is given by:
\begin{equation}
\sigma_{\rm v} \sim {{\rm St}^2 \over 1 + \alpha \, {\rm St}^2}\,,
\label{N46}
\end{equation}
where $\alpha \sim 1$.

\section{The particle velocity field for ${\rm St} \gg 1$}
\label{s:crit-val}

The equation of motion of a particle reads:
\begin{equation}
\frac{d{\mathbf{v}}(t,{\mathbf{r}})}{dt} ={1 \over \tau_p}
[{\mathbf{u}}(t, {\mathbf{r}})-{\mathbf{v}}(t,{\mathbf{r}})]\;,
\label{39}
\end{equation}
where the total time derivative $(d/dt) $ takes into account the
time dependence of the particle coordinate $\mathbf{r} $:
\begin{equation}
\frac{d}{dt}=\left[\frac{\partial}{\partial
t}+{\mathbf{v}}(t,{\mathbf{r}})\cdot\nabla\right]\ . \label{40}
\end{equation}
Now Eq.~(\ref{39}) takes the form:
\begin{equation}
\left\{\tau_p\left[\frac{\partial}{\partial
t}+{\mathbf{v}}(t,{\mathbf{r}})\cdot\nabla\right]+1\right\}
{\mathbf{v}}(t,{\mathbf{r}})={\mathbf{u}}(t,{\mathbf{r}}) \ .
\label{41}
\end{equation}
In the following we analyze this equation for particles with the
time $\tau_p$ which is larger than the turnover time of the smallest
eddies in the Kolmogorov micro-scale $\tau(\eta)$, but is smaller
than the turnover time of the largest eddies $\tau(L)$. Denote by
$\ell_*$ the characteristic scale of eddies for which
\begin{equation}\label{eq:def-l*}
\tau_p= \tau(\ell_*)\ .
\end{equation}
This scale as well as the particle cluster scale was introduced in
\cite{EKB96}. Note that $\ell_* / \eta = {\rm St}^{3/2}$. The eddies
with $\ell \gg \ell_*$ almost fully involve particles in their
motions, while the eddies with $\ell \ll \ell_*$ do not affect the
particle motions in the zero order approximation with respect to the
ratio $[\tau(\ell)/\tau_p]\ll 1 $. Therefore it is conceivable to
suggest that the main contribution to the particle velocity is due
to the eddies with the scale of $\ell$ [which we denote as $\B
v_\ell (t,\r)$] that is of the order of $\ell_*$ and much larger
then the Kolmogorov micro-scale. Velocity $\B v_\ell (t,\r)$ cannot
be found on the basis of simple dimensional reasoning because the
problem at hand involves a number of dimensionless parameters like
$\ell/\ell_*$, $\ell_*/\eta$, etc. The main difficultly in
determining this velocity is that in this case one has to take into
account for a modification of the particle response time $\tau_p$ by
the turbulent fluctuations. The physical reason for that is quite
obvious: the time $\tau_p$ is determined by molecular viscosity of
the carrier fluid while the main dissipative effect for motions with
$\ell>\eta$ is due to the effective ``turbulent" viscosity. In order
to determine the velocity $\B v_\ell (t,\r)$ we can use the
perturbation approach to Eq.~(\ref{41}) (see, e.g.,
\cite{LPA95,LPB95}). The details of this derivations are given in
Appendix A. This analysis yields the scalings of the particle
velocity for ${\rm St} \gg 1$:
\begin{equation}
{v}^{2}_{\ell}\approx u^{2}_{\ell} \left(\frac{\ell}
{\ell_*}\right)^{10/9} \approx u^{2}_{\ell}
\left[\frac{\tau(\ell)} {\tau_p}\right]^{5/3} \ .\label{65}
\end{equation}

\section{The clustering instability of the second moment
of particle number density} \label{moments}

In this section we will perform a quantitative analysis for the
clustering instability of the second moment of particle number
density. To determine the growth rate of the clustering instability
let us consider the equation for the two-point correlation function
$\Phi(t, \B R)$ of particle number density:
\begin{eqnarray}
{\partial \Phi \over \partial t} = [B(\B R) + 2 {\B U}(\B R)\cdot
\B {\nabla} + \hat D_{\alpha \beta}(\B R) \nabla_{\alpha}
\nabla_{\beta}] \, \Phi(t,\B R) \,, \label{WW6}
\end{eqnarray}
(see \cite{EK02}). The meaning of the coefficients $ B(\B R) $, $\B
U(\B R) $ and $ \hat D_{\alpha \beta}(\B R)$ is as follows (for
details see Appendix B). The function $ B(\B R) $ is determined by
the compressibility of the particle velocity field and it causes the
generation of fluctuations of the number density of particles. The
vector $ \B U(\B R) $ determines a scale-dependent drift velocity
which describes a transport of fluctuations of particle number
density from smaller scales to larger scales, i.e., in the regions
with larger turbulent diffusion. The latter can decrease the growth
rate of the clustering instability. Note that $ {\bf U}(\B R=0) = 0
$ whereas $ B(\B R=0) \not= 0 .$ For incompressible velocity field $
{\bf U}(\B R) = 0 $ and $B(\B R) = 0$. The scale-dependent tensor of
turbulent diffusion $ \hat{D}_{\alpha\beta}(\B R)$ is also affected
by the compressibility. In very small scales this tensor is equal to
the tensor of the molecular (Brownian) diffusion, while in the
vicinity of the maximum scale of turbulent motions this tensor
coincides with the regular tensor of turbulent diffusion.

Thus, the clustering instability is determined by the competition
between these three processes. The form of the coefficients $ B(\B
R) $, $\B U(\B R) $ and $ \hat D_{\alpha \beta}(\B R)$ depends on
the model of turbulent velocity field. For instance, for the random
velocity with Gaussian statistics of the particle trajectories these
coefficients are given in Appendix B.

Let us study the clustering instability. We consider particles with
the size $\eta / \sqrt{\rm Sc} \ll a \ll \eta $, where ${\rm Sc}=
\nu / D$ is the Schmidt number. For small inertial particles
advected by air flow ${\rm Sc} \gg 1$. There are three
characteristic ranges of scales, where the form of the solution of
Eq.~(\ref{WW6}) for the two-point correlation function $\Phi(t, \B
R)$ of the particle number density is different. These ranges of
scales are the following: (i) the dissipative range $a \leq \ell
\leq \eta$, where the molecular diffusion term $\propto 1/{\rm Sc}$
is negligible; (ii) the first part of the inertial range $\eta \leq
\ell \leq \ell_\ast$ and (iii) the second part of the inertial range
$\ell_\ast \ll \ell \ll L$, where the functions $ B(\B R) $ and $\B
U(\B R) $ are negligibly small.

Consider a solution of Eq.~(\ref{WW6}) in the vicinity of the
thresholds of the excitation of the clustering instability. The
asymptotic solution of the equation for the two-point correlation
function $\Phi(t, \B R)$ of the particle number density is obtained
in Appendix B. In the range of scales $a \leq \ell \leq \eta$, the
correlation function $\Phi(t, \B R)$ in a non-dimensional form reads
\begin{eqnarray} \label{L8}
\Phi(R) = A_{1} R^{-\lambda_d} \sin( \mu_d |\ln \, R| + \varphi_d)
\,,
\end{eqnarray}
and in the range of scales $\eta \leq \ell \leq \ell_\ast$ it is
given by
\begin{eqnarray} \label{RL8}
\Phi(R) = A_{2} R^{-\lambda} \sin( \mu \ln \, R + \varphi) \,,
\end{eqnarray}
where the parameters  $\lambda_d$ and $\mu_d$ are given by
Eq.~(\ref{ML15}) and the parameters $\lambda$ and $\mu$ are given by
Eq.~(\ref{MML15}) in Appendix B. Here $R$ is measured in the units
of $\eta$ and time $t$ is measured in the units of $\tau_\eta \equiv
\tau(\ell=\eta)$. We have taken into account that the correlation
function $\Phi(R)$ has a global maximum at $ R = a$, i.e. the
normalized correlation function of the particle number density
$\Phi(t, R=a) = 1$. We have also taken into account that in the
range of scales $\eta \leq \ell \ll \ell_\ast$, the relationship
between $v^{2}_{\ell}$ and $u^{2}_{\ell}$ is given by:
\begin{eqnarray}
v^{2}_{\ell} = u^{2}_{\ell} \left[\frac{\tau(\ell)}
{\tau_p}\right]^{s} \ . \label{L1}
\end{eqnarray}
For instance, for ${\rm St} \gg 1$ the exponent $s=5/3$ (see
Eq.~(\ref{65})). The value $s = 7/4$ corresponds to the turbulent
diffusion tensor with the scaling $\propto R^2$ [see
Eqs.~(\ref{RW12})-(\ref{RW15}) in Appendix B]. We consider the
parameter $s$ as a phenomenological parameter. In the range of
scales $\ell_\ast \ll \ell \ll L$, the correlation function $\Phi(t,
\B R)$ is given by
\begin{eqnarray}\label{L7}
\Phi(R) = A_{3} R^{-\lambda_3} \,,
\end{eqnarray}
where $\lambda_3$ is given by Eq.~(\ref{LL7}) in Appendix B. The
condition, $\int_{0}^{\infty} R^{2} \Phi(R) \,d R = 0$, implies that
the total number of particles in a closed volume is conserved.

The growth rate of the second moment of particle number density, the
coefficients $A_k$ and the parameters $\varphi_d$, $\varphi$ are
determined by matching the correlation function $ \Phi(R) $ and its
first derivative $ \Phi'(R) $ at the boundary of the above three
ranges of scales, i.e., at the points $\ell =\eta$ and $\ell =
\ell_\ast .$ For example, the growth rate $\gamma$ of the clustering
instability of the second-order correlation function is given by
\begin{eqnarray}
\gamma &=& {1 \over 6 \, \tau_\eta (1 + 3 \sigma_{_{T}})} \biggl[400
\, \sigma_{\rm v} \, {\sigma_{_{T}} - \sigma_{\rm v} \over 1 +
\sigma_{\rm v}} - {(3 - \sigma_{_{T}})^{2} \over 1 + \sigma_{_{T}}}
\nonumber \\
&& - 4 \mu_d^2 \,  {(1 + 3 \sigma_{_{T}})^{2} \over 1 +
\sigma_{_{T}}} \biggr] \;, \label{WG6}
\end{eqnarray}
where $\sigma_{_{T}}$ is the degree of compressibility of the
scale-dependent tensor of turbulent diffusion
$\hat{D}_{\alpha\beta}(\B R)$ (for details, see Appendix B). Note
that for the $ \delta $-correlated in time random Gaussian
compressible velocity field, the coefficients $ B(\B R) $ and $\B
U(\B R) $ are related to the turbulent diffusion tensor $ \hat
D_{\alpha \beta}(\B R)$, i.e.,
\begin{eqnarray}
B(\B R) = \nabla_{\alpha} \nabla_{\beta} \hat D_{\alpha \beta}(\B R)
\,,  \quad U_{\alpha}(\B R) = \nabla_{\beta} \hat D_{\alpha
\beta}(\B R) \,, \label{WLL6}
\end{eqnarray}
(for details, see \cite{EKA00,EKA01,EK02}). In this case the second
moment $ \Phi(t,\B R) $ can only decay, in spite of the
compressibility of the velocity field. For the $ \delta $-correlated
in time random Gaussian compressible velocity field $\sigma_{\rm
v}=\sigma_{_{T}}$. For the finite correlation time of the turbulent
velocity field $\sigma_{_{T}} \not=\sigma_{\rm v}$ and the
relationships~(\ref{WLL6}) are not valid. The clustering instability
depends on the ratio $\sigma_{_{T}} / \sigma_{\rm v}$.

The range of parameters $ (\sigma_{\rm v} , \sigma_{_{T}})$ for
which the clustering instability of the second moment of particle
number density may occur is shown in Fig.~1. The line $ \sigma_{\rm
v} = \sigma_{_{T}} $ corresponds to the $ \delta $-correlated in
time random compressible velocity field for which the clustering
instability cannot be excited. The various curves indicate results
for different value of the parameter $s$. The curves for $s = 7/4$
(dashed) and $s=5/3$ (solid) practically coincide. The parameter $s$
is considered as a phenomenological parameter, and the change of
this parameter from $s=7/4$ to $s=0$ can describe a transition from
one asymptotic behaviour (in the range of scales $\eta \leq \ell
\leq \ell_\ast$) to the other ($\ell_\ast \ll \ell \leq L$). The
growth rate~(\ref{WG6}) of the clustering instability versus
$\sigma_{\rm v}$ for $s=5/3$ and different values of $\sigma_{_{T}}$
is shown in Fig.~2.

\begin{figure}
\centering
\includegraphics[width=8cm]{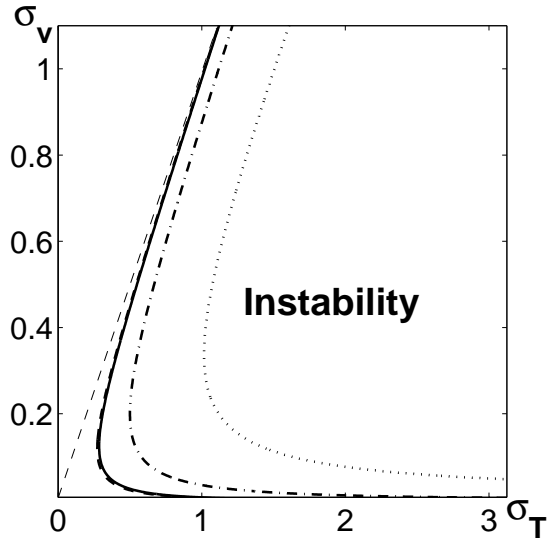}
\caption{\label{Fig1} The range of parameters $ (\sigma_{\rm v} ,
\sigma_{_{T}})$ for which the clustering instability may occur. The
various curves indicate results for $s = 7/4$ (dashed), $s=5/3$
(solid), for $s=1$ (dashed-dotted) and for $s=2/3$ (dotted). The
thin dashed line $ \sigma_{\rm v} = \sigma_{_{T}} $ corresponds to
the $ \delta $-correlated in time random compressible velocity
field.}
\end{figure}

\begin{figure}
\centering
\includegraphics[width=8cm]{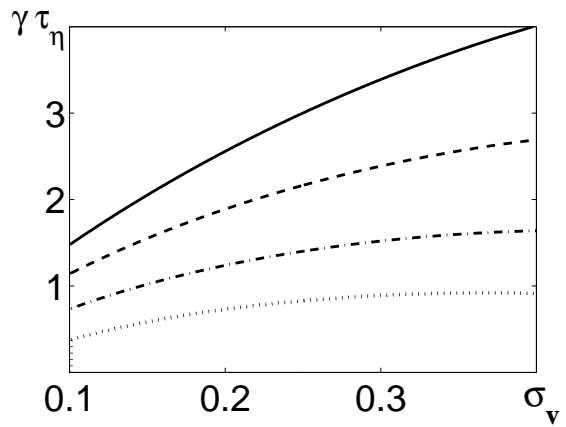}
\caption{\label{Fig2} The growth rate of the clustering instability
versus $\sigma_{\rm v}$ for $s=5/3$. The various curves indicate
results for different values of $\sigma_{_{T}} = \sigma_{\rm v} +
\delta$: $\; \delta=3$ (solid), $\, \delta=1$ (dashed), $\delta=0.5$
(dashed-dotted) and $\delta=0.3$ (dotted).}
\end{figure}

We have not discussed in the present study the growth of the
high-order moments of particle number density (see
\cite{EKB96,EK02,BF01,ZM88}). The growth of the negative moments of
particles number density (possibly associated with formation of
voids and cellular structures) was discussed in
\cite{BF01,SZ89,KS97}.

\section{Discussion}
\label{s:disc}

Formation and evolution of particle clusters are of fundamental
significance in many areas of environmental sciences, physics of the
atmosphere and meteorology [smog and fog formation, rain formation
(see e.g., \cite{S86,FS88,PK97,KS01,CK04,CC05}), planetary physics
(see e.g., \cite{HB98,BC99}), transport and mixing in industrial
turbulent flows, like spray drying and cyclone dust separation,
dynamics of fuel droplets (see e.g., \cite{CS98,H88,BR99}). The
analysis of the experimental data showed that the spatial
distributions of droplets in clouds are strongly inhomogeneous (see
\cite{S03}). The small-scale inhomogeneities in particle
distribution were observed also in laboratory turbulent flows (see
\cite{FK94,AC02,AG06}).

In the present study we have shown that the particle spatial
distribution in the turbulent flow field is unstable against
formation of clusters with particle number density that is much
higher than the average particle number density. Obviously this
exponential growth at the linear stage of instability should be
saturated by nonlinear effects. A momentum coupling of particles and
turbulent fluid is essential when the kinetic energy of fluid $ \rho
\langle {\B u}^{2} \rangle $ is of the order of the particles
kinetic energy $ m_{\rm p} n_{\rm cl} \langle {\B v}^{2} \rangle ,$
where $ |{\B u}| \sim |{\B v}|$, i.e., when $ m_{\rm p} n_{\rm cl}
\sim \rho$. This condition implies that the number density of
particles  in the cluster $n_{\rm cl} \sim a^{-3} (\rho/3 \rho_{\rm
p})$. In the atmospheric turbulence the characteristic parameters
are as follows: in the viscous scale, $\eta\sim 1$ mm, the
correlation time of the turbulent velocity field is $\tau_\eta \sim
(0.01 - 0.1)$ s, and for water droplets $ \rho_{\rm p} / \rho \sim
10^{3}$. Thus, for $ a \sim 30 \, \mu$m we obtain $ n_{\rm cl} \sim
10^{4} $ cm$^{-3}$ (see \cite{EK02}). Particle collisions can play
also essential role when during the life-time of a cluster the total
number of collisions is of the order of number of particles in the
cluster. The collisions in clusters may be essential for $n_{\rm cl}
\sim a^{-3} (\ell_\ast/a)  (\rho/3 \rho_{\rm p})$. In this case a
mean separation of particles in the cluster is of the order of
$\ell_{\rm s} \sim a^{4/3} (3 \rho_{\rm p}/\ell_\ast \rho)^{1/3}$.
When, e.g., $a \sim 30 \, \mu$m we get $ \ell_{\rm s} \sim 5
\,a\approx 150 \, \mu$m and $n_{\rm cl} \sim 3\times
(10^{4}-10^{5})$ cm$^{-3}$. The mean number density of droplets in
clouds $N$ is about $10^3$ cm$^{-3}$. Therefore, the clustering
instability of droplets in clouds can increase their concentrations
in the clusters by the order of magnitude (see also \cite{EK02}).
Note that for large Stokes numbers the terminal fall velocity of
particles can be much larger than the turbulent velocity. This
implies that the sedimentation of heavy particles can suppress the
clustering instability for large Stokes numbers.

\begin{figure}
\centering
\includegraphics[width=8cm]{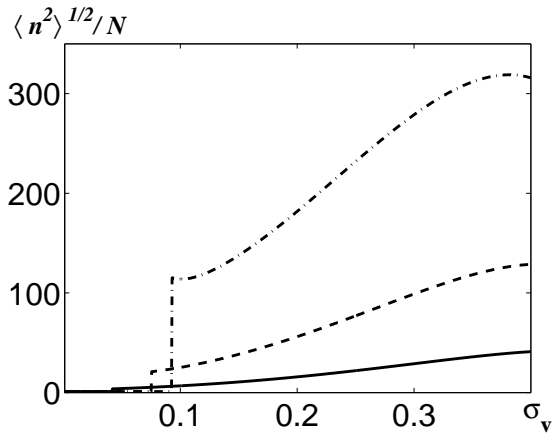}
\caption{\label{Fig3} Dependence of $\sqrt{\langle n^2 \rangle} / N$
versus parameter $\sigma_{\rm v}$ for $s=5/3$ which corresponds to
the solution of Eq.~(\ref{WW6}) for the two-point correlation
function $\Phi(t, \B R)$ of the particle number density obtained in
Sec. IV. Various curves indicate results for different values of the
parameter $\delta$: $\; \delta=3$ (solid), $\, \delta=1$ (dashed)
and $\delta=0.5$ (dashed-dotted), where $\sigma_{_{T}} = \sigma_{\rm
v} + \delta$.}
\end{figure}

\begin{figure}
\centering
\includegraphics[width=8cm]{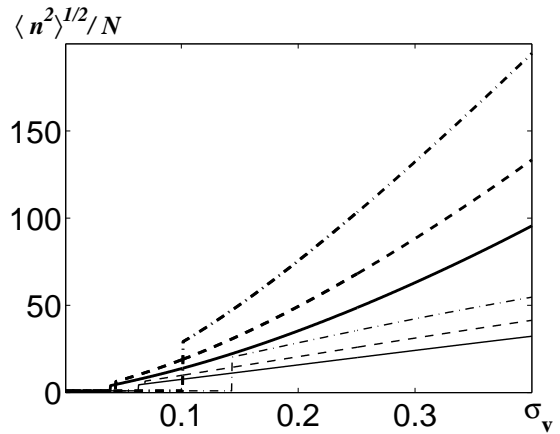}
\caption{\label{Fig4} Dependence of $\sqrt{\langle n^2 \rangle} / N$
versus parameter $\sigma_{\rm v}$ which corresponds to the solution
for the two-point correlation function $\Phi(t, \B R)$ of the
particle number density for the case ${\rm St} < 1$ studied in
\cite{EK02}. Various curves indicate results for different values of
the parameter $\delta$: $\; \delta=3$ (solid), $\, \delta=1$
(dashed) and $\delta=0.3$ (dashed-dotted), where $\sigma_{_{T}} =
\sigma_{\rm v} + \delta$. The thick lines correspond to ${\rm
Sc}=10^5$ and thin lines correspond to ${\rm Sc}=10^4$.}
\end{figure}

There is an additional restriction on the value of $n_{\rm cl} =
\sqrt{\langle n^2 \rangle}$ which follows from the condition
$\langle n(t, \mathbf{r}) \, n(t, \mathbf{r}+\mathbf{R}) \rangle
\equiv N^2 + \langle \Theta^2 \rangle \Phi(t, \mathbf{R}) \geq 0$,
where $\Phi(t, \mathbf{R})$ is the normalized correlation function
of the particle number density $\Phi(t, R=a) = 1$. Since the
correlation function $\Phi(t, \mathbf{R})$ can be negative at some
scale $R$, this condition implies that the maximum possible value of
$\langle \Theta^2\rangle$ which can be achieved during the
clustering instability is $\langle \Theta^2 \rangle_{\rm max} = N^2
/ |\Phi|_{\rm min}$. Therefore, the number density of particles in
the cluster $n_{\rm cl}$ cannot be larger than $n_{\rm cl} \leq N (1
+ |\Phi|_{\rm min}^{-1})^{1/2}$. Using this criterion we plotted in
Fig.~3 the dependencies $\sqrt{\langle n^2 \rangle}_{\rm max} / N$
versus parameter $\sigma_{\rm v}$ for different values of the
parameter $\delta$, where $\sigma_{_{T}} = \sigma_{\rm v} + \delta$.
We estimate $|\Phi|_{\rm min}$ using the solution of Eq.~(\ref{WW6})
for the two-point correlation function $\Phi(t, \B R)$ of the
particle number density obtained in Sec. IV. Note however, that this
solution determines the linear stage of the clustering instability.
In Fig.~3 we also take into account the conditions for the
clustering instability. This condition implies that for a given
parameter $\sigma_{_{T}}$ the clustering instability is excited when
$\sigma_{\rm v} > \sigma_{\rm v}^{\rm min}$ (see Fig.~1). For
comparison we also plotted in Fig.~4 the similar dependencies
$\sqrt{\langle n^2 \rangle}_{\rm max} / N$ versus parameter
$\sigma_{\rm v}$  using the solution for the two-point correlation
function $\Phi(t, \B R)$ of the particle number density for the case
${\rm St} < 1$ studied in \cite{EK02}.

In the present study we have considered the particle clustering due
to the clustering instability. Generally, particle clustering can
also occur due to the source of fluctuations of droplets number
density $I = B({\bf R}) N^2$ in Eq.~(\ref{WW6}) for the second-order
correlation function of particle number density. This source term
arises due to the term $\propto - N \, {\rm div}\,{\bf v}$ in
Eq.~(\ref{6}). Such fluctuations were studied in
\cite{B03,BL04,BF01}.

Note that there is an alternative approach which determines the
particle clustering (see \cite{DM05,MW05,WM05}). The particle number
density fluctuations are generated by a multiplicative random
process: volume elements in the particle flow are randomly
compressed or expanded, and the ratio of the final density to the
initial density after many multiples of the correlation time $\tau$
can be modelled as a product of a large number of random factors.
According to this picture, the particle number density fluctuations
will be a record of the history of the flow, and may bear no
relation to the instantaneous disposition of vortices when the
particle number density is measured \cite{DM05,MW05,WM05}. The
particle number density is expected to have a log-normal probability
distribution. When the random-flow model \cite{DM05,MW05,WM05} with
short correlation time is applied to fully-developed turbulence it
predicts that the clustering is strongest when ${\rm St} \sim 1$, in
agreement with numerical studies \cite{CK04,CC05}.

\section{Conclusions}
\label{s:concl}

In this study we considered formation of small-scale clusters of
inertial particles in a turbulent flow. The mechanism for particle
clustering is associated with a small-scale instability of particle
spatial distribution. The clustering instability is caused by a
combined effect of the particle inertia and a finite correlation
time of the turbulent velocity field. The theory of particle
clustering developed in our previous studies was extended to the
case when the particle Stokes time is larger than the Kolmogorov
time scale, but is much smaller than the correlation time at the
integral scale of turbulence. We found the criterion for the
clustering instability for this case.

\bigskip

\leftline{\bf Acknowledgments}

This research was supported in part by The German-Israeli Project
Cooperation (DIP) administrated by the Federal Ministry of Education
and Research (BMBF), by the Israel Science Foundation governed by
the Israeli Academy of Science, by Binational Israel - United States
Science Foundation (BSF), by the Israeli Universities Budget
Planning Committee (VATAT) and Israeli Atomic Energy Commission, by
Swedish Ministry of Industry (Energimyndigheten, contract P
12503-1), by the Swedish Royal Academy of Sciences, the STINT
Fellowship program.

\appendix

\section{Velocity of  inertial particles
for $\tau_p \gg \tau_\eta$}

\renewcommand{\theequation}
            {A.\arabic{equation}}

In order to determine the velocity $\B v_\ell (t,\r)$ we can use the
Wyld's perturbation diagrammatic approach to Eq.~(\ref{41}) in the
Belinicher-L'vov representation (see, e.g., \cite{LPA95,LPB95}).
This approach yields automatically a sensible result allowing us to
avoid an overestimation  of the sweeping effect in an order-by-order
perturbation analysis. However, keeping in mind that this approach
is technically quite involved, in this study we reformulated the
derivation procedure and obtained the required results using a more
simple procedure based on the equation of motion Eq.~(\ref{41}).

To determine ${\bm{v}}_{\ell}(t,{\bm{r}}) $ we consider
Eq.~(\ref{41}) in the  frame moving with $\ell$-eddies, in which the
surrounding fluid velocity $\B u$ equals to the relative velocity of
the $\ell$-eddy at $\B r$, i.e., $\B u \tr = \B u_\ell\tr$. Here one
has to take into account that the $\ell$-eddy is swept out by all
eddies with scales $ \ell' > \ell$. At the same time the particles
participate in motions of $\ell'$-eddies with $ \ell' > \ell_* >
\ell$. Therefore, the relative velocity $ \B v_{\ell}$ of the
$\ell$-eddy and the particle is determined by $\ell'$-eddies with
the intermediate scales, $\ell_*\leq  \ell' \leq \ell$. This
velocity is determined by the contribution of $\ell_*$-eddies, and
can be considered as a time and space independent constant $\B u_*$
during the life time of the $\ell$ eddy and inside it. Velocity $\B
u_*$ in our approach is random and has the same statistics as the
statistics of  the turbulent velocities of $\ell_*$-eddies. Then
Eq.~(\ref{41}) becomes
\begin{eqnarray}
&&\left(\tau_p\frac{\partial}{\partial t} +
1\right){\bm{v}}_{\ell}(t,{\bm{r}}) = {\bm{u}}_{\ell}(t,{\bm{r}+\B
u_* t})
\nonumber\\
&& \quad \quad - \tau_p\left[{\bm{v}}_{\ell}
(t,{\bm{r}})\cdot\nabla\right] {\bm{v}}_{\ell}(t,{\bm{r}})\ .
\label{eq:linn}
\end{eqnarray}
In Eq.~(\ref{eq:linn}) the velocity $\B u_\ell$ is calculated at
point $\B r$ and the velocity $\B v_\ell$ is at $\B r - \B u_*\, t$.
For the sake of convenience we redefine here $\B r - \B u_*\, t= \B
r' $ as $ \B r $ and, respectively, $\B r = \B r'+\B u_*\, t $ as $
\B r+\B u_*\, t$. Note that Eq.~(\ref{eq:linn}) is a simplified
version of Eq.~(\ref{41}) that we used in our derivations.

\subsection{First non-vanishing contribution to $v_\ell$}

Since $ v_\ell \tr \ll u_\ell$ for $\ell \ll \ell_*$, we can find
the first non-vanishing contribution to
${\bm{v}}_{\ell}(t,{\bm{r}})$ in the limit $[\tau(\ell)\ll \tau_p]$
by considering the linear version of Eq.~(\ref{eq:linn}):
\begin{equation}
\left(\tau_p\frac{\partial}{\partial
t}+1\right){\bm{v}}_{\ell}(t,{\bm{r}})
={\bm{u}}_{\ell}(t,{\bm{r}}+\B u_* t) \ . \label{eq:v-u-linear}
\end{equation}
In the $\omega, \B k$ representation this equation takes the form:
\begin{equation}
\left(i \omega\, \tau_p +1\right){\bm{v}}_{\ell}\ok
={\bm{u}}_{\ell}(\omega - \B k \cdot \B u_* ,\B k) \,,
\label{eq:v-u-lin-ko}
\end{equation}
that allows one to find the relationship between the second order
correlation functions $F_{v,\ell}^{\alpha\beta}\ok$ and
$F_{u,\ell}^{\alpha\beta}\ok$ of the velocity fields $\B v_\ell$
and $\B u_\ell$:
\begin{equation}\label{eq:Fv-Fu}
F_{v,\ell}^{\alpha\beta}\ok= \frac1{\omega^2 \tau_p^2+1}
F_{u,\ell}^{\alpha\beta}(\omega - \B k \cdot \B u_* ,\B k)\ .
\end{equation}
Functions $F_{u,\ell}^{\alpha\beta}\ok$ and $F_{v,\ell} ^{\alpha
\beta} \ok$ are defined as usual, e.g.,
\begin{eqnarray}\label{eq:def-Fu}
&&(2 \pi)^4\delta(\omega+\omega')\, \delta (\B k + \B
k')F_{u,\ell}^{\alpha\beta}\ok \equiv
\nonumber\\
&& \< v_\ell^\alpha \ok\, v_\ell^\beta (\omega',\B k')\> \; .
\end{eqnarray}

The simultaneous correlation functions are related to their
$\omega$-dependent counterparts via the integral $\int d \omega/
2\pi$, e.g.,
\begin{equation}\label{eq:rel-1}
F_{v,\ell}^{\alpha\beta}(\B k)=\int \frac{d\omega}{2\pi}
F_{v,\ell}^{\alpha\beta}\ok\ .
\end{equation}
The tensorial structure of $ F_{u,\ell} ^{\alpha\beta}(\B k)$
follows from the incompressibility condition and the assumption of
isotropy:
\begin{equation}\label{eq:incom}
F_{u,\ell} ^{\alpha\beta}(\B k)=P^{\alpha\beta}(\B k)
F_{u,\ell}(k)\,,
\end{equation}
where $P^{\alpha\beta}(\B k)$ is the transversal projector:
\begin{equation}\label{eq:trans}
P^{\alpha\beta}(\B k)=\delta_{\alpha\beta}-k^\alpha k^\beta/k^2\ .
\end{equation}
In the inertial range of scales the function $F_{u,\ell}
^{\alpha\beta}\ok$ may be written in the following form:
\begin{equation}\label{eq:Fu-1}
F_{u,\ell} ^{\alpha\beta}\ok = P^{\alpha\beta}(\B k)\,
F_{u,\ell}(k) \, \tau (\ell)\, f [\omega\, \tau (\ell)]\ .
\end{equation}
Here the dimensionless function $f(x)$ is normalized as follows:
\begin{equation}\label{eq:norm}
\int_{-\infty}^\infty f(x)\, dx = 2\pi\ .
\end{equation}

Now we can average Eq.~(\ref{eq:Fv-Fu}) over the statistics of
$\ell_*$-eddies. Denoting the mean value of some function $g(x)$
as $\overline{g(x)}$ we have:
\begin{eqnarray}\label{eq:ev-1}
&& \overline{f [(\omega - \B k \cdot \B u_*) \,\tau (\ell)]} \approx
\overline{\delta [(\omega- \B k \cdot \B u_*)\, \tau(\ell)]}
\nonumber\\
&& \approx \frac \ell{\tau(\ell)\, u_*} \, f_* \left(\frac {u_*
\,\omega}\ell\right)\ .
\end{eqnarray}
Here the dimensionless function $f_*(x)$ has one maximum at $x=0$,
and it is normalized according to Eq.~(\ref{eq:norm}). The
particular form of $f_*(x)$ depends on the statistics of
$\ell_*$-eddies and our qualitative analysis is not sensitive to
this form. Thus, we may choose, for instance:
\begin{equation}\label{eq:Lor}
f_*(x)=2 /[x^2+1]\ .
\end{equation}
In Eqs.~(\ref{eq:ev-1}) we took into account that the
characteristic Doppler frequency of $\ell$-eddies (in the random
velocity field $u_*$ of $\ell_*$-eddies) may be evaluated as:
\begin{equation}\label{eq:D-1}
\gamma_{_{D}}(\ell)\equiv \sqrt{\overline{( \B k \cdot \B u_*)^2
}} \simeq u_*/\ell\ .
\end{equation}
This frequency is much larger than the characteristic frequency
width of the function $f[\omega \tau(\ell)]$ (equal to
$1/\tau(\ell)$), and therefore the function $f(x)$ in
Eq.~(\ref{eq:ev-1}) may be approximated by the delta function
$\delta(x)$.

After averaging, Eq.~(\ref{eq:Fv-Fu}) may be written as
\begin{equation}\label{eq:Fv}
F_{v,\ell}^{\alpha\beta}\ok= \frac{P^{\alpha\beta}(\B k)
f_*(0)}{\omega^2 \tau_p^2+1}\frac \ell{u_*} F_{u,\ell}(k)\ .
\end{equation}
Here we took into account that $\tau_p\gg \ell/u_*$ that allows us
to neglect the frequency dependence of $ f_*(u_* \,\omega/\ell)$
and to calculate this function at $\omega=0$. Together with
Eq.~(\ref{eq:Fv}) this yields
\begin{equation}\label{eq:rel-2}
F_{v,\ell}^{\alpha\beta}(\B k)=P^{\alpha\beta}(\B k)\,
F_{u,\ell}(k)\frac { \,\ell} {\, \tau_p\, u_*}\,,
\end{equation}
where we used the estimate $ f_*(0)\approx 2$, that follows from
Eq.~(\ref{eq:Lor}).

The equation~(\ref{eq:rel-2}) provides the relationship between
the mean square relative velocity of $\ell$-separated particles,
$v_\ell$, and the velocity of $\ell$-eddies, $u_\ell$:
\begin{eqnarray}\label{eq:v-u-1}
v_\ell\simeq u_\ell \sqrt{\frac { \,\ell} {\, \tau_p\, u_*}}
\simeq u_\ell \sqrt{\frac { \,\ell} {\ell _*}}\ .
\end{eqnarray}

\subsection{Effective nonlinear equation}

For a qualitative analysis of the role of the nonlinearity of the
particle behavior in the $\ell $-cluster we evaluate $\nabla $ in
the nonlinear term, Eq.~(\ref{eq:linn}), as $1/\ell $, neglecting
the spatial dependence and the vector structure. The resulting
equation in $\omega $-representation reads:
\begin{eqnarray}\nonumber
(i\omega&+&\gamma_p)V_{\ell}(\omega)=\gamma_ {\rm
dr}U_{\ell}(\omega)+{\mathcal{N}}_{\omega}\,, \quad
\gamma_p=1/\tau_p\, , \\
{\mathcal{N}}_{\omega}&=&-\frac{1}{2\pi \ell}\int
d\omega_1d\omega_{2}
\delta(\omega+\omega_1+\omega_{2})V_{\ell}(\omega_1)\label{eq:nonlin}
V_{\ell}(\omega_{2})\, ,\nonumber \\
V_{\ell}(\omega)&=&\int v_{\ell}(t)\exp[-i\omega t]dt\,,
\nonumber \\
v_{\ell}(t)&=&\frac{1}{2\pi}\int V_{\ell}(\omega)\exp[i\omega
t]d\omega\ .
\end{eqnarray}
In the zeroth order (linear) approximation
$({\mathcal{N}}_{\omega}\rightarrow 0) $
\begin{equation}
V_{\ell}^{(0)}(\omega)=\frac{\gamma_p U_{\ell}(\omega)}
{i\omega+\gamma_p}\,,\label{52}
\end{equation}
which is the simplified version of Eq.~(\ref{eq:v-u-lin-ko}). This
allows us to find in the linear approximation
\begin{equation}
\langle v_{\ell}^{2}(t)\rangle =\int\frac{d\omega}{2\pi}
{\mathcal{F}}_{\ell} (\omega)=\int\frac{d\omega}{2\pi}
\frac{\gamma_p^{2}\overline{F_{u,\ell}(\omega)}}
{\omega^{2}+\gamma_p^{2}}\, , \label{53}
\end{equation}
where $F_{u,\ell}(\omega)$ is the correlation functions of
$U_{\ell}(\omega) $:
\begin{equation}
2\pi\delta(\omega+{\omega}^{\prime})F_{u,\ell} (\omega)=\langle
U_{\ell} (\omega)U_{\ell}({\omega}^{\prime})\rangle\,, \label{54}
\end{equation}
similarly to Eq.~(\ref{eq:def-Fu}).

In the limit $\tau_p\gg \ell/u_* $ one can neglect in
Eq.~(\ref{53}) the $\omega $-dependence of
$\overline{F_{u,\ell}(\omega)} $, which has the characteristic
width $\ell/u_* $ and conclude:
\begin{eqnarray} \nonumber
v^{2}_{\ell,0}&\equiv&\langle v_{\ell}^{2}(t)\rangle
\approx\frac{\gamma_p}{2} \overline{F_{u,\ell}(0)}\approx
u^{2}_{\ell} \frac{\ell\, \gamma_p} {u_*}\approx u^{2}_{\ell}
\frac\ell{\ell_*},
\nonumber \\
u^{2}_{\ell}&\equiv&\langle u_{\ell}^{2}(t)\rangle\,, \label{55}
\end{eqnarray}
in agreement with Eq.~(\ref{eq:v-u-1}).

\subsection{First nonlinear correction}

To evaluate the first nonlinear correction to Eq.~(\ref{55}) one
has to substitute $V_{\ell} (\omega) $ from Eq.~(\ref{52}) into
Eq.~(\ref{eq:nonlin}) for ${\mathcal{N}}_{\omega}$:
\begin{eqnarray}\nonumber
V_{\ell,1}(\omega)&=& -\frac{\gamma^{2}_p}{2\pi \ell} \int
d\omega_1d\omega_{2}\delta(\omega+\omega_1+\omega_{2})
\frac{U_{\ell}(\omega_1)}{i\omega_1+\gamma_p}
\nonumber\\
 && \times\frac{U_{\ell}(\omega_{2})}
{i\omega_{2}+\gamma_p}\ .\label{56}
\end{eqnarray}
Using Eq.~(\ref{56}) instead of Eq.~(\ref{52}) we obtain instead
of Eq.~(\ref{53})
\begin{eqnarray}\label{57}
v^{2}_{\ell,1}&\equiv& \langle [v_{\ell,1}(t)]^2 \rangle =
\int\frac{d\omega}{2\pi}{F_{u,\ell ,1}} (\omega)\, ,\\
F_{u,\ell ,1} (\omega)& = & \frac{2\gamma^{4}_p}
{(\omega^{2}+\gamma_p^{2})\ell^{2}}
\int\frac{d\omega_1d\omega_{2}}{2\pi} \frac{\delta(\omega+\omega_1
+\omega_{2})}{(\omega^{2}_1+\gamma_p^{2})
(\omega^{2}_{2}+\gamma_p^{2})} \nonumber\\
 && \times
\overline{F_{u,\ell} (\omega_1)} \ \overline{F_{u,\ell}(\omega_{2})}
\ . \nonumber
\end{eqnarray}
In this derivation we assumed for simplicity the Gaussian statistics
of the velocity field. This corresponds to a standard closure
procedure in theory of turbulence (see, e.g., \cite{MY75,MC90}).
Taking into account of deviations from the Gaussian statistics of
the turbulent velocity field in the framework of the perturbation
theory of turbulence does not yield qualitatively new results due to
the general structure of the series in the theory of perturbations
after Dyson-Wyld line-resummation (see, e.g., \cite{LPA95,LPB95}).

Now let us estimate
\begin{equation}
v^{2}_{\ell,1} \approx\frac{\left[\overline{F_{u,\ell}(0)}\right
]^{2}}{\ell^{2}}\approx \frac {u^{4}_{\ell}}{u_*^2 }\approx
u^{2}_{\ell}\left( \frac{\ell}{\ell_*}\right)^{2/3}\, , \label{58}
\end{equation}
that is much larger than the result (\ref{55}) for $v^{2}_{\ell,0}
$ obtained in the linear approximation. This means that the simple
iteration procedure we used is inconsistent, since it involves
expansion in large parameter [$(\ell_*/\ell)^{1/3}$].

\subsection{Renormalized perturbative expansion}

A similar situation with a perturbative expansion occurs in the
theory of hydrodynamic turbulence, where a simple iteration of the
nonlinear term with respect to the linear (viscous) term, yields the
power series expansion in ${\mathcal{R}}e^{2}\gg 1 $. A way out,
used in the theory of hydrodynamic turbulence is the Dyson-Wyld
re-summation of one-eddy irreducible diagrams (for details see,
e.g., \cite{LPA95,LPB95,LPC95}). This procedure corresponds to
accounting for the nonlinear (so-called "turbulent" viscosity)
instead of the molecular kinematic viscosity. A similar approach in
our problem implies that we have to account for the self-consistent,
nonlinear renormalization of the particle frequency
$\gamma_p\Rightarrow\Gamma_p(\ell) $ in Eq.~(\ref{eq:nonlin}) and to
subtract the corresponding terms from
$\tilde{{\mathcal{N}}}_{\omega} $. With these corrections,
Eq.~(\ref{eq:nonlin}) reads:
\begin{eqnarray}
[\,i\omega+\Gamma_p(\ell)\,]V_{\ell}(\omega)=\gamma_p U_{\ell}
(\omega)+\tilde{{\mathcal{N}}}_{\omega}\ . \label{eq:Ncor}
\end{eqnarray}
Here $\tilde{{\mathcal{N}}}_{\omega} $ is the nonlinear term
${\mathcal{N}}_{\omega} $ after substraction of the nonlinear
contribution to the difference
\begin{equation}
\Delta_p\equiv\Gamma_p(\ell)-\gamma_p
\approx\frac{v_{\ell}^{2}/\ell^{2}}{\Gamma_p(\ell)} \ . \label{60}
\end{equation}
The latter relation actually follows from a more detailed
perturbation diagrammatic approach. In our context it is
sufficient to realize that in the limit $\Gamma_p(\ell)\gg\gamma_p
$ one may evaluate $\Gamma_p(\ell) $ by a simple dimensional
reasoning:
\begin{equation}
\Gamma_p(\ell)\approx v_{\ell}/\ell\, ,\label{61}
\end{equation}
which is consistent with Eq.~(\ref{60}). In addition,
Eq.~(\ref{60}) has a natural limiting case
$\Gamma_p(\ell)\rightarrow\gamma_p $ when
$v_{\ell}/\ell\ll\gamma_p $. Now using Eq.~(\ref{eq:Ncor}) instead
of Eq.~(\ref{52}) we arrive at:
\begin{equation}
V_{\ell}^{(0)}(\omega)=\frac{\gamma_p U_{\ell}
(\omega)}{i\omega+\Gamma_p(\ell)}\ .\label{62}
\end{equation}
Accordingly, instead of the estimates~(\ref{55}) one has:
\begin{equation}
{v}^{2}_{\ell,0}\approx u^{2}_{\ell}\,\frac{\gamma_p^{2}\, \ell}
{\Gamma_p(\ell)\,u_* }\approx \,u^{2}_{\ell}\,
\frac{\gamma_p}{\Gamma_p(\ell)} \left(\frac{\ell}{\ell_*}\right)\,
. \label{63}
\end{equation}
The latter equation together with Eq.~(\ref{61}) allows to
evaluate $\Gamma_p(\ell)$ as follows:
\begin{equation}
\Gamma_p(\ell)\approx\left (\frac{\gamma_p^{2} u_\ell^2}{ \ell\,
u_* }\right )^{1/3}\approx
\gamma_p\left(\frac{\ell_*}{\ell}\right)^{1/9} \ .\label{64}
\end{equation}
Hence the estimate~(\ref{63}) becomes
\begin{equation}
{v}^{2}_{\ell,0}\approx u^{2}_{\ell} \left(\frac{\ell}
{\ell_*}\right)^{10/9} \approx u^{2}_{\ell}
\left[\frac{\tau(\ell)} {\tau_p}\right]^{5/3} \ .\label{65A}
\end{equation}
Repeating the evaluation of the nonlinear correction
${v}^{2}_{\ell,2} $ with the renormalized Eq.~(\ref{eq:Ncor}) we
find that
\begin{equation}
{v}^{2}_{\ell,1} \approx {v}^{2}_{\ell,0}\ . \label{66}
\end{equation}
This means that now the expansion parameter is of the order of 1, in
accordance with  the renormalized perturbation approach. This
procedure yields Eq.~(\ref{65}).

\section{The clustering instability of the inertial particles}

\renewcommand{\theequation}
            {B.\arabic{equation}}

The clustering instability is determined by the equation for the
two-point correlation function $\Phi(t, \B R)$ of particle number
density (see Eq.~(\ref{WW6})). The tensor $\hat{D}_{\alpha\beta}(\B
R)$ may be written as
\begin{eqnarray}
\hat{D}_{\alpha\beta}(\B R) &=& 2 D \delta_{\alpha\beta} +
D^{^{T}}_{\alpha \beta}(\B R)\,,
 \nonumber\\
 D_{\alpha \beta}^{^{T}} (\B R) &=& \tilde{D}_{\alpha \beta}^{^{T}} (0)
- \tilde{D}_{\alpha \beta}^{^{T}} (\B R)\ .
\end{eqnarray}
The form of the coefficients $ B(\B R) $, $\B U(\B R) $ and $ \hat
D_{\alpha \beta}(\B R)$ in Eq.~(\ref{WW6}) depends on the model of
turbulent velocity field. For instance, for the random velocity with
Gaussian statistics of the Wiener trajectories $\B \xi(t,\B r|\tau)$
these coefficients are given by
\begin{eqnarray}
\label{LL30} B(\B R) &\approx& 2 \int_{0}^{\infty} \langle b[0,\B
\xi(t,\B r_1|0)] b[\tau,\B \xi(t,\B r_2|\tau)] \rangle \,d \tau\,,
 \br
{\B U}(\B R) & \approx & - 2 \int_{0}^{\infty} \langle {\B
v}[0,\B \xi(t,\B r_1|0)] b[\tau,\B \xi(t,\B r_2|\tau)] \rangle
\,d\tau \,,
 \br
\tilde D_{\alpha \beta}^{^{T}} (\B R) &\approx& 2
\int_{0}^{\infty} \langle v_{\alpha}[0,\B \xi(t,\B r_1|0)]
v_{\beta}[\tau,\B \xi(t,\B r_2|\tau)] \rangle \,d \tau \,,
\end{eqnarray}
where $ b={\rm div} \, {\mathbf{v}} $ (for more details, see
\cite{EK02}). Note that in this study we use Eulerian description.
In particular, in Eq.~(\ref{LL30}) the functions $v_\alpha[\tau,
\bec{\xi}(t,{\bf r}|\tau)]$ and $b[\tau,\bec{\xi}(t,{\bf r}|\tau)]$
describe the Eulerian velocity and its divergence calculated at the
Wiener trajectory (see \cite{EK02,EKB00,EKB01}). The Wiener
trajectory $\bec{\xi}(t,{\bf r}|s)$ (which usually is called the
Wiener path) and the Wiener displacement $\bec{\rho}_w(t,{\bf r}|s)$
are defined as follows:
\begin{eqnarray*}
\bec{\xi}(t,{\bf r}|s) &=&{\bf r} - \bec{\rho}_w(t,{\bf r}|s)\,,
\nonumber\\
\bec{\rho}_w(t,{\bf r}|s) &=& \int^{t}_{s} {\bf v}
[\tau,\bec{\xi}(t,{\bf r}|\tau)] \,\,d \tau + \sqrt{2 D} \, {\bf
w}(t-s) \,,
\end{eqnarray*}
where ${\bf w}(t)$ is the Wiener random process which describes the
Brownian motion (molecular diffusion). The Wiener random process
${\bf w}(t)$ is defined by the following properties: $\langle {\bf
w}(t) \rangle_{\bf w}=0\,, $ $\, \langle w_i(t+\tau) w_j(t)
\rangle_{\bf w}= \tau \delta _{ij}$, and $ \langle \dots
\rangle_{\bf w} $ denotes the mathematical expectation over the
statistics of the Wiener process. Since $v_\alpha[\tau,
\bec{\xi}(t,{\bf r}|\tau)]$ describes the Eulerian velocity
calculated at the Wiener trajectory, the Wiener displacement
$\bec{\rho}_w(t,{\bf r}|s)$ can be considered as an Eulerian field.
We calculate the divergence of the Eulerian field of the Wiener
displacements $\bec{\rho}_w(t,{\bf r}|s)$.

Now we introduce the parameter $\sigma_{_{T}}$ which is defined by
analogy with Eq.~(\ref{19}):
\begin{eqnarray}
\label{S} \hskip -0.4cm \sigma_{_{T}}\equiv \frac{\B \nabla \cdot \B
D_{_{T}}\cdot \B \nabla }{\B \nabla \times \B D_{_{T}}\times \B
\nabla }= \frac{\nabla_\alpha \nabla_\beta D^{^{\rm T}}_{\alpha
\beta}(\B R)}{\nabla_\alpha\nabla_\beta D^{^{\rm T}} _{\alpha'
\beta'}(\B R) \epsilon_{\alpha\alpha'\gamma}
\epsilon_{\beta\beta'\gamma} }\,,
\end{eqnarray}
where $ \epsilon_{\alpha\beta\gamma}$ is the fully antisymmetric
unit tensor. Equations~(\ref{19}) and~(\ref{S}) imply that $
\sigma_{_{T}}=\sigma_{\rm v}$ in the case of $\delta$-correlated in
time compressible velocity field.

For a random incompressible velocity field with a finite correlation
time the tensor of turbulent diffusion $ \tilde D^{^{\rm T}}_{\alpha
\beta} (\B R) = \tau^{-1} \langle \xi_\alpha(t,\B r_1|0)
\xi_\beta(t,\B r_2|\tau) \rangle $ (see \cite{EK02}) and the degree
of compressibility of this tensor is
\begin{eqnarray}
\label{SN} \sigma_{_{T}} = {\langle (\B \nabla \cdot \B {\xi})^{2}
\rangle \over \langle (\B \nabla {\bf \times} \B {\xi})^{2} \rangle}
\, .
\end{eqnarray}

Let us study the clustering instability. We consider particles of
the size $\eta / \sqrt{\rm Sc} \ll a \ll \eta $, where ${\rm Sc}=
\nu / D $ is the Schmidt number. For small inertial particles
advected by air flow ${\rm Sc} \gg 1 $. A general form of the
turbulent diffusion tensor in a dissipative range is given by
\begin{eqnarray} \label{RW12}
&& D^{^{\rm T}}_{\alpha \beta}(\B R)=  C_{1}^d R^{2} \delta_{\alpha
\beta} + C_{2}^d R_\alpha R_\beta  \,,
\\  \nonumber
&&C_{1}^d = {2 (2 + \sigma_{_{T}}) \over 3\, (1 + \sigma_{_{T}})}\,,
\quad  C_{2}^d = {2 (2\sigma_{_{T}} - 1) \over 3\,( 1 +
\sigma_{_{T}})}\; .
\end{eqnarray}
In the range of scales $a \leq \ell \leq \eta$, Eq.~(\ref{WW6}) in a
non-dimensional form reads:
\begin{eqnarray}
{\partial \Phi \over \partial t} = R^2 \Phi'' (C_{1}^d + C_{2}^d) +
2\, R \Phi' (U_d + C_{1}^d) + B_d \Phi \,, \label{RW14}
\end{eqnarray}
where $R$ is measured in the units of $\eta$, time $t$ is measured
in the units of $\tau_\eta \equiv \tau(\ell=\eta)$, and the
molecular diffusion term $\propto 1/{\rm Sc}$ is negligible.
Consider a solution of Eq.~(\ref{RW14}) in the vicinity of the
thresholds of the excitation of the clustering instability. Thus,
the solution of~(\ref{RW14}) in this region is
\begin{eqnarray}
\Phi(r) = A_{1} R ^{-\lambda_1} \,, \label{RW15}
\end{eqnarray}
where $\lambda_1 = \lambda_d \pm i \mu_d$ and
\begin{eqnarray}
\lambda_d &=& {C_{1}^d - C_{2}^d + 2 U_d \over 2 (C_{1}^d +
C_{2}^d)} \;, \quad \mu_d = {C_3^d \over 2 (C_{1}^d + C_{2}^d)} \,,
\label{ML15} \\
(C_3^d)^2 &=&4 (B_d -\gamma) \, (C_1^d +C_2^d) - (C_{1}^d - C_{2}^d
+ 2 U_d)^{2} \ , \nonumber
\end{eqnarray}
and
\begin{eqnarray*}
B_d = 20 \, {\sigma_{\rm v} \over \sigma_{\rm v} + 1} \,, \quad U_d
= (1/3) \, B_d \, .
\end{eqnarray*}
Since the correlation function $ \Phi(R) $ has a global maximum at $
R = a ,$ the coefficient $ \, C_{1}^d > C_{2}^d - 2U_d$ if $\mu_d$
is a real number (see below).

Consider the range of scales $\eta \leq \ell \ll \ell_\ast$. The
relationship between $v^{2}_{\ell}$ and $u^{2}_{\ell}$ is determined
by Eq.~(\ref{L1}), where according to Eq.~(\ref{65}) the exponent
$s=5/3 .$ In this case the expression for the turbulent diffusion
tensor in non-dimensional form reads
\begin{eqnarray} \label{W12}
&& D^{^{\rm T}}_{\alpha \beta}(\B R)= R^{(4s-7)/3} (C_{1} R^{2}
\delta_{\alpha \beta} + C_{2} R_\alpha R_\beta ) \,,
\\ \nonumber
&&C_{1} = { 5 + 4 s + 6 \sigma_{_{T}} \over 9 \, (1 +
\sigma_{_{T}})}\,, \quad C_{2} = {(4 s - 1) (2\sigma_{_{T}} - 1)
\over 9 \,( 1 + \sigma_{_{T}})} \, .
\end{eqnarray}

To determine the functions $B(\B R)$ and $\B U(\B R)$ in the range
of scales $\eta \leq \ell \ll \ell_\ast$ we use the general form of
the two-point correlation function of the particle velocity field in
this range of scales:
\begin{eqnarray*}
&& \langle v_{\alpha}(t,\B r) v_{\beta}(t+\tau,\B r + \B R) \rangle
= {1 \over 3} [\delta_{\alpha \beta} - (C_{1}^v R^{2} \delta_{\alpha
\beta}
\nonumber\\
&&+ \, C_{2}^v R_\alpha R_\beta)\, R^{2(s-2)/3}] f(\tau) \,,
\nonumber\\
&&C_{1}^v = { (4 + s + 3 \sigma_{\rm v})\over 3 \, (1 + \sigma_{\rm
v})}\,, \quad C_{2}^v = {(1 + s) (2\sigma_{\rm v} - 1) \over 3 \,( 1
+ \sigma_{\rm v})} \, .
\end{eqnarray*}
Substitution this equation into Eq.~(\ref{LL30}) yields
\begin{eqnarray}
\label{L2} \B U(\B R) = U_0 \, R^{(4s-7)/3} \,, \quad B(\B R) = B_0
\, R^{(4s-7)/3} \,,
\end{eqnarray}
where
\begin{eqnarray*}
U_0 = \beta_1 \, {\sigma_{\rm v} \over \sigma_{\rm v} + 1} \,, \quad
B_0 = \beta_2 \, U_0 \,
\end{eqnarray*}
and the coefficients $\beta_1$ and $\beta_2$ depend on the
properties of turbulent velocity field.  The dimensionless functions
$B_0$ and $U_0$  in Eq.~(\ref{L2}) are measured in the units of
$\tau_\eta^{-1}$.

For the $ \delta $-correlated in time random Gaussian compressible
velocity field $\sigma_{_{T}}=\sigma_{\rm v}$ (for details, see
\cite{EKA00,EKA01,EK02}). In this case the second moment $ \Phi(t,\B
R) $ can only decay, in spite of the compressibility of the velocity
field. For the finite correlation time of the turbulent velocity
field $\sigma_{_{T}} \not=\sigma_{\rm v}$ and Eqs.~(\ref{WLL6}) are
not valid. The clustering instability depends on the ratio
$\sigma_{_{T}} / \sigma_{\rm v}$. In order to provide the correct
asymptotic behaviour of Eq.~(\ref{L2}) in the limiting case of the $
\delta $-correlated in time random Gaussian compressible velocity
field, we have to choose the coefficients $\beta_1$ and $\beta_2$ in
the form:
\begin{eqnarray*}
\beta_1 = 8 (4 s^2 + 7 s - 2) / 27 \,, \, \, \beta_2 = (4 s + 2 ) /
3 \, .
\end{eqnarray*}
Note that when $s< 1/4$, the parameters $\beta_1 < 0 $ and $B(\B R)
< 0$. In this case there is no clustering instability of the second
moment of particle number density. Thus, Eq.~(\ref{WW6}) in a
non-dimensional form reads:
\begin{eqnarray}
{\partial \Phi \over \partial t} &=& R^{(4s-7)/3} [R^2 \Phi'' (C_{1}
+ C_{2}) + 2\, R \Phi' (U_0 + C_{1})
\nonumber\\
&& + B_0 \Phi] \, . \label{W14}
\end{eqnarray}
Consider a solution of Eq.~(\ref{W14}) in the vicinity of the
thresholds of the excitation of the clustering instability, where
$(\partial \Phi / \partial t) R^{(7-4s)/3}$ is very small. Thus, the
solution of~(\ref{W14}) in this region is
\begin{eqnarray}\label{L6}
\Phi(R) = A_{2} R^{-\lambda_2} \,,
\end{eqnarray}
where $ \lambda_2 = \lambda \pm i \mu ,$
\begin{eqnarray}\label{MML15}
\lambda &=& {C_{1} - C_{2} + 2 U_0 \over 2 (C_{1} + C_{2})} \,,
\quad \mu = {C_3 \over 2 (C_{1} + C_{2})} \,, \br C_3^2&=&4
B_0(C_1+C_2) - (C_{1} - C_{2} + 2 U_0)^{2} \ .
\end{eqnarray}
Since the correlation function $ \Phi(R) $ has a global maximum at $
R = a ,$ the coefficient $ \, C_{1} > C_{2}- 2U_0$ if $\mu$ is a
real number (see below).

Consider the range of scales $\ell_\ast \ll \ell \ll L$. In this
case the non-dimensional form of the turbulent diffusion tensor is
given by
\begin{eqnarray} \label{L4}
&& D^{^{\rm T}}_{\alpha \beta}(\B R)= R^{-2 / 3} (\tilde C_{1} R^{2}
\delta_{\alpha \beta} + \tilde C_{2} R_\alpha R_\beta) \,,
\\ \nonumber
&&\tilde C_{1} = { 2 (5 + 3 \sigma_{_{T}})\over 9 \, (1 +
\sigma_{_{T}})}\,, \; \; \tilde C_{2} = {4 (2 \sigma_{_{T}} - 1)
\over 9 \,( 1 + \sigma_{_{T}})} \,,
\end{eqnarray}
and Eq.~(\ref{WW6}) reads:
\begin{eqnarray}
{\partial \Phi \over \partial t} &=& R^{-2 / 3} [R^2 \Phi'' (\tilde
C_{1} + \tilde C_{2}) + 2\, R \Phi' \tilde C_{1}] \, . \label{L5}
\end{eqnarray}
Here we took into account that in this range of scales the functions
$ B(\B R) $ and $\B U(\B R) $ are negligibly small. Consider a
solution of Eq.~(\ref{L5}) in the vicinity of the thresholds of the
excitation of the clustering instability, when $(\partial \Phi /
\partial t) R^{2/3}$ is very small. The solution of~(\ref{L5}) is
given by
\begin{eqnarray}\label{LLL7}
\Phi(R) = A_{3} R^{-\lambda_3} \,,
\end{eqnarray}
where
\begin{eqnarray}\label{LL7}
\lambda_3 = {|\tilde C_{1} - \tilde C_{2}| \over \tilde C_{1} +
\tilde C_{2}} = {|7 - \sigma_{_{T}}| \over 3 + 7 \sigma_{_{T}}} \, .
\end{eqnarray}

The growth rate of the second moment of particle number density can
be obtained by matching the correlation function $ \Phi(R) $ and its
first derivative $ \Phi'(R) $ at the boundary of the above three
ranges of scales, i.e., at the points $\ell =\eta$ and $\ell =
\ell_\ast .$ Such matching is possible only when $ \lambda_2 $ is a
complex number, i.e., when $ C_3^2 > 0$ (i.e., $\mu$ is a real
number). The latter determines the necessary condition for the
clustering instability of particle spatial distribution. It follows
from Fig. 1 that in the range of parameters where $\mu$ is a real
number, the parameter $\mu_d$ is also a real number. The asymptotic
solution of the equation for the two-point correlation function
$\Phi(t, \B R)$ of the particle number density in the range of
scales $a \leq \ell \leq \eta$  is given by
Eqs.~(\ref{L8})-(\ref{RL8}).

\end{document}